\documentclass{article} 
 
\usepackage{amsmath,amssymb,amsthm,latexsym} 
 
\usepackage{stmaryrd,wasysym,upgreek,mathrsfs,dsfont} 
\usepackage[english]{babel} 
\usepackage{graphicx,color} 

\newtheorem{lemma}{Lemma}[section]
\newtheorem{theorem}{Theorem}[section]
\newcommand{\beq}{\begin{equation}}
\newcommand{\eeq}{\end{equation}}
\newcommand{\beqa}{\begin{eqnarray}}
\newcommand{\eeqa}{\end{eqnarray}}

\newcommand{\si}{\sigma}

\newcommand{\Tr}{{\rm Tr}}

\begin{document} 
 \title{Constructive $\phi^4$ field theory without tears} 
\author{J. Magnen$^{1}$, V. Rivasseau$^{2}$\\
1) Centre de Physique Th\'eorique, CNRS UMR 7644,\\
Ecole Polytechnique F-91128 Palaiseau Cedex, France\\
2) Laboratoire de Physique Th\'eorique, CNRS UMR 8627,\\ 
Universit\'e Paris XI,  F-91405 Orsay Cedex, France}

\maketitle 
\begin{abstract} 
We propose to treat the $\phi^4$ Euclidean theory
constructively in a simpler way. Our method, based
on a new kind of "loop vertex expansion",
no longer requires the painful intermediate tool of cluster and Mayer expansions.
\end{abstract} 
 
\section{Introduction} 

Constructive field theory build functions whose Taylor expansion 
is perturbative field theory \cite{GJ,Riv1}. Any formal power series
being asymptotic to infinitely many smooth functions, perturbative field theory alone
does not give any well defined mathematical recipe to compute to arbitrary accuracy
any physical number, so in a deep sense it is no theory at all.

In field theory ``thermodynamic" or infinite volume quantities
are expressed by connected functions. One main advantage of perturbative field theory 
is that connected functions are simply the sum of the connected Feynman graphs.
But the expansion diverges because there are too many such graphs.
However to know connectedness does not require the full knowledge of a Feynman graph 
(with all its loop structure) but only the (classical) notion of a spanning tree in it. 
This remark is at the core of the developments of constructive field theory, 
such as cluster expansions, summarized in the constructive golden rule:

\emph{``Thou shall not know most of the loops, or thou shall diverge!"}

Some time ago Fermionic constructive theory was quite radically simplified. It was realized
that it is possible to rearrange perturbation theory \emph{order by order} by grouping together
pieces of Feynman graphs which share a common tree \cite{Les,AR2}.
This is made easily with the help of a universal combinatoric so-called
forest formula \cite{BK,AR1} which once and for all essentially solves
the problem that a graph can have many spanning trees. Indeed
it splits any amplitude of any connected graph in a certain number of pieces and
attributes them in a "democratic" and "positivity preserving" way 
between all its spanning trees. Of course the possibility for such a rearrangement
to lead to convergent resummation of Fermionic perturbation
theory ultimately stems from the Pauli principle which is responsible 
for \emph{analyticity} of that expansion in the coupling constant. 

Using this formalism Fermionic theory can now 
be manipulated at the constructive level almost as easily as at 
the "perturbative level to all orders".
It lead to powerful mathematical physics theorems such as for instance those about 
the behavior of interacting Fermions in 2 dimensions \cite{DR1,FKT,Hub}, and to 
more explicit constructions \cite{DR2}
of just renormalizable Fermionic field theories such as the Gross-Neveu model in 
two dimensions first built in \cite{GK,FMRS}. 

But bosonic constructive theory remained awfully difficult.
To compute the thermodynamic functions, until today one needed to introduce two
different expansions one of top of the other. 
The first one, based on a discretization of space into a lattice of cubes
which breaks the natural rotation invariance of the theory, is called a cluster expansion.
The result is a dilute lattice gas of clusters but 
with a remaining hardcore interaction. Then a second expansion called Mayer expansion
removes the hardcore interaction. The same tree formula is used {\textit {twice}}
once for the cluster and once for the Mayer
expansion\footnote{It is possible  to combine both expansions into a single one \cite{AMR},
but the result cannot be considered a true simplification.}, the breaking of 
rotation invariance to compute rotation invariant quantities seems \emph{ad hoc}
and the generalization of this technique to many renormalization group steps
is considered so difficult that despite courageous attempts
towards a better, more explicit formalization \cite{Br,AR3}, it remains until now confined 
to a small circle of experts. 

The bosonic constructive theory cannot be simply rearranged in a convergent series
\emph{order by order} as in the Fermionic case, because all graphs at a given order
have the same sign. Perturbation theory has zero convergence radius for bosons.
The oscillation which allows resummation (but only e.g. in the Borel sense) of the perturbation
theory must take place between infinite families of graphs of different orders. 
To explicitly identify such families and rearrange the perturbation theory accordingly
seemed until now very difficult. The cluster and Mayer expansion perform this task but in a very 
complicated and indirect way.

In this paper we at last identify such infinite families of graphs. They 
give rise to an explicit convergent expansion for the connected functions
of bosonic $\phi^4$ theory, without any lattice and cluster or Mayer expansion.
In fact we stumbled upon this new method by trying to adapt former cluster expansions to 
large matrix $\phi^4$ models in order to extend constructive methods to non-commutative field theory
(see \cite{Riv2} for a recent review). The matrix version is described in a separate 
publication \cite{Riv3}. Hopefully it should allow a non-perturbative construction 
of the $\phi^{\star 4}$ theory on Moyal space ${\mathbb R}^4$, whose 
renormalizable version was pioneered by Grosse and Wulkenhaar \cite{GW}.

\section{The example of the pressure of $\phi^4$}

We take as first example the construction of the pressure
of $\phi^4_4$ in a renormalization group (RG) slice. The goal is e.g. to prove its Borel summability
in the coupling constant uniformly in the slice index,
without using any lattice (breaking Euclidean invariance)
nor any cluster or Mayer expansion. 

The propagator in a RG slice $j$ is e.g.
\begin{equation}\label{bound}
C_j (x,y) = \int^{M^{-2j +2}}_{M^{-2j}} e^{-\alpha m^2} 
e^{- (x-y)^2/4\alpha }{\alpha^{-2}}d\alpha \le KM^{2j} e^{-c M^{j} \vert x-y \vert}
\end{equation}
where $M$ is a constant defining the size of the RG slices, and $K$ and $c$ from now on are generic 
names for inessential constants, respectively large and small. We could also use compact support cutoffs
in momentum space to define the RG slices.

Consider a local interaction $\lambda \int  \phi^4 (x) d^4x =\lambda\Tr  \phi^4 $
where the trace means spatial integration. For the moment assume the coupling $\lambda$ 
to be real positive and small. We decompose the $\phi^4$ functional integral according to 
an intermediate field as:
\begin{equation}\label{intermediate}
\int d\mu_{C_j}(\phi) e^{-  \lambda\Tr \phi^4 } = \int d\nu(\si) 
e^{-\frac 12 \Tr \log (1 + i H) }
\end{equation}
where $d\nu$ is the ultralocal measure on $\sigma$ with covariance $\delta(x-y)$,
and $H= \lambda^{1/2}  D_j \sigma   D_j $ is an Hermitian operator, 
with $D_j = C_j^{1/2}$.

The pressure is known to be the Borel sum
of all the connected vacuum graphs with a particular 
root vertex fixed at the origin. We want to prove this through a new method.

\begin{figure}[!htb]
\centering
\includegraphics[scale=0.5]{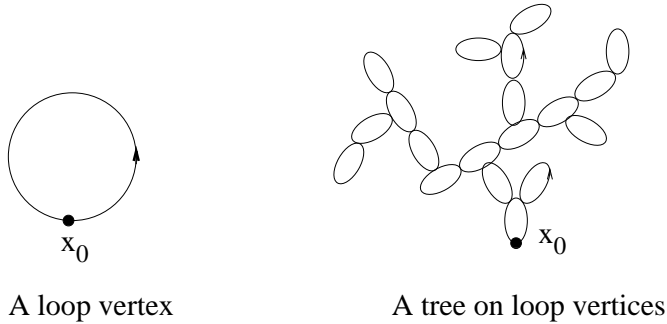}
\caption{Loop vertices and a tree on them}
\label{looptree}
\end{figure}

We define the \emph{loop vertex}\footnote{To avoid any confusion with the former $\phi^4$
vertices we shall not omit the word \emph{loop}.}
$V=- \frac 12 \Tr \log (1 + i H )  $. This loop vertex can be pictured
as in the left hand side of Figure \ref{looptree}. The trace means integration
over a ``root" $x_0$. Cyclic invariance means that this root can be moved everywhere
over the loop. It is also convenient to also introduce an arrow, 
by convention always turning counterclockwise
for a $+iH$ convention, and anti-clockwise for a complex conjugate loop vertex
$\bar V=- \frac 12 \Tr \log (1 - i H )  $.

We then expand the exponential as $\sum_n \frac{V^n}{n!} $. To compute the connected
graphs we give a (fictitious) index $v$, $v=1,..., n$ to all the $\sigma$ fields of a given 
loop vertex $V_v$. This means that we consider 
$n$ different copies $\sigma_v$ of $\sigma$ with a 
degenerate Gaussian measure $d\nu (\{\sigma_v\})$ whose
covariance is $<\sigma_v \sigma_{v'}>_{\nu} = \delta(x-y)$. The functional
integral over $d\nu (\sigma)$ is equal to the  functional
integral over $d\nu (\{\sigma_v\})$.
We apply then the forest formula of \cite{AR1} to test connexions between the 
loop vertices from 1 to $n$. (The lines of this forest, which join loop vertices 
correspond to former $\phi^4$ vertices.) 

The logarithm of the partition function $\log Z(\Lambda)$
at finite volume $\Lambda$ is given
by this formula restricted to trees (like in the Fermionic case \cite{AR2}), and 
spatial integration restricted to $\Lambda$. The pressure or 
infinite volume limit of $\frac {\log Z(\Lambda)} {\vert \Lambda\vert }$ is given 
by the same \emph{rooted} tree formula but with one particular position fixed at the origin, 
for instance the position associated to a particular root line $\ell_0$. More precisely:

\begin{theorem}

\begin{eqnarray}\label{treeformul}
\lim_{\Lambda \to {\mathbb R}^4}\frac {\log Z(\Lambda)} {\vert \Lambda\vert } 
&=& \sum_{n=1}^{\infty} \frac{1}{n!}\sum_T        \bigg\{ \prod_{\ell\in T}   
\big[ \int_0^1 dw_\ell   \big]\bigg\} G_T(\sigma, x_{\ell_0})\vert_{x_{\ell_0} =0}  \\
G_T(\sigma, x_{\ell_0})&=&\prod_{\ell\in T}   \int d^4 x_\ell d^4 y_\ell 
\int  d\nu_T (\{\sigma_v\}, \{ w \})  \nonumber \\
\hskip-1cm && \bigg\{ \prod_{\ell\in T} \big[ \delta (x_\ell - y_\ell) 
 \frac{\delta}{\delta \sigma_{v(\ell)}(x_\ell)}\frac{\delta}{\delta \sigma_{v'(\ell)}(y_\ell)} 
\big] \bigg\} \prod_v V_v , \label{gt}
\end{eqnarray}
where 
\begin{itemize}

\item each line $\ell$ of the tree joins two different vertices $V_{v(\ell)}$ and $V_{v'(\ell)}$ 
at point $x_{\ell}$ and $y_{\ell}$, which are identified through the function
$\delta (x_\ell - y_\ell) $ (since the covariance of $\sigma$ is ultralocal),

\item the sum is over rooted trees over $n$ vertices, which have therefore
$n-1$ lines, with root $\ell_0$, 

\item the normalized Gaussian measure $d\nu_T (\{\sigma_v\}, \{ w \})  $ over the vector field $\sigma_v$ has covariance
$$<\sigma_v,\sigma_{v'}>=
\delta (x-y) w^T (v, v', \{ w\})$$ where $w^T (v, v', \{ w\})$ is 1 if $v=v'$,
and the infimum of the $w_\ell$ for $\ell$ running over the unique path from $v$ to $v'$ in $T$
if $v\ne v'$. This measure is well-defined because the matrix $w^T$ is positive.

\end{itemize}
\end{theorem}

This is indeed the outcome of the universal tree formula of \cite{AR1} in this case. 
To check it, we need only to move by cyclicity the local root of each loop 
nearest to the global root in the tree. This global root 
point is chosen for simplicity in formulas above at a particular root line $\ell_0$,
but in fact it could be fixed anywhere in an arbitrarily chosen ``root loop",
as shown on the right hand side of Figure \ref{looptree} 
(with all loops oriented counterclockwise).
 
\begin{figure}[!htb]
\centering
\includegraphics[scale=0.5]{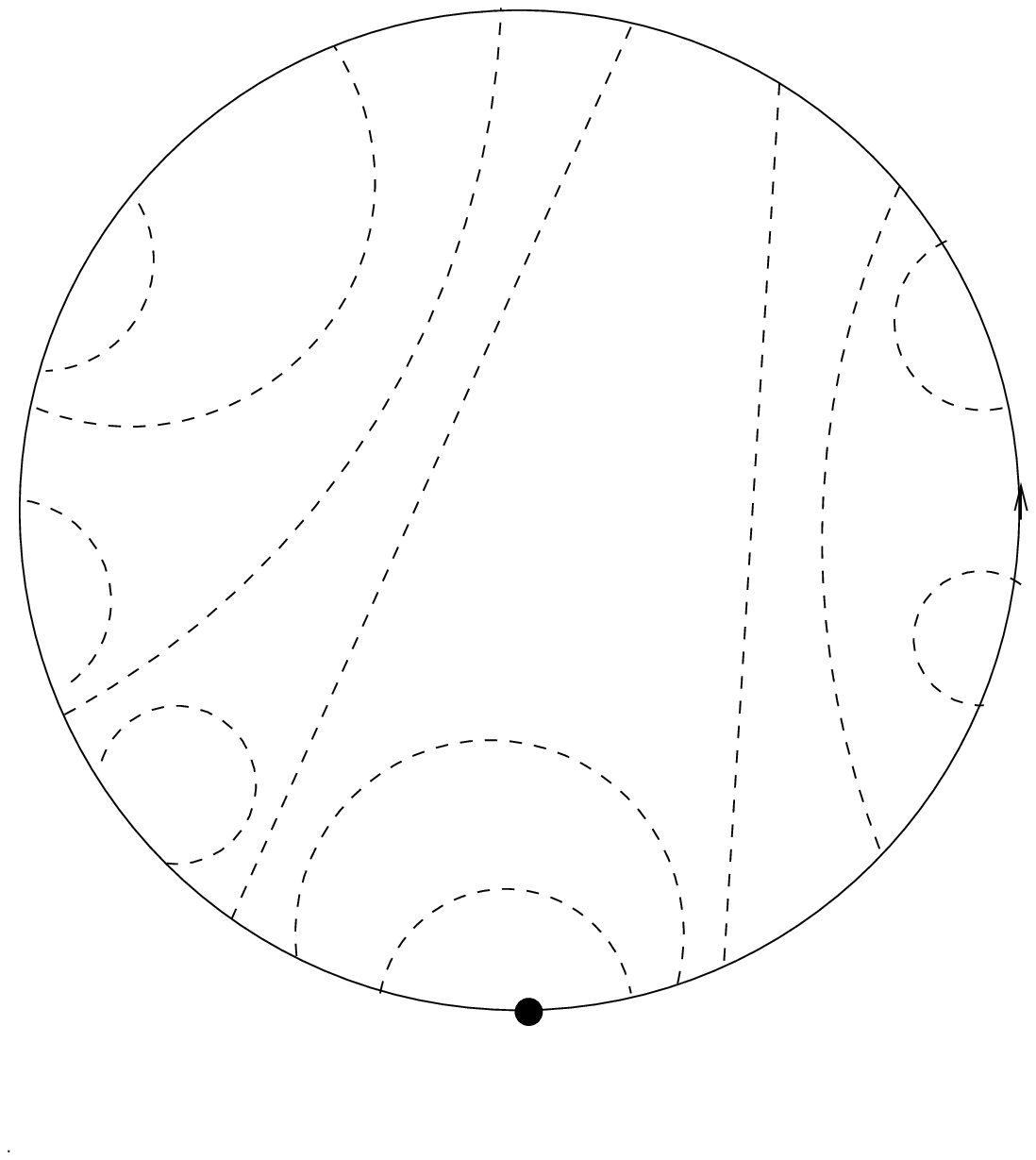}
\caption{The big loop representation}
\label{bigloop}
\end{figure}

But there is  an other representation of the same object.
A tree on connecting loops such as the one shown in the right hand side of 
Figure \ref{looptree} can also be drawn as a set of dotted lines dividing 
in a  \emph{planar} way
a \emph{single loop} as in Figure \ref{bigloop}. 
Each dotted line carries a $\delta (x_\ell - y_\ell) $ function 
which identifies pairs of points on the border of the loop joined by the dotted line, 
and is equipped with a coupling constant,
because it corresponds to an old $\phi^4$ vertex. 
This second picture is obtained by turning around the tree.
The pressure corresponds to the sum over such planar partitions
of a single big loop with an arbitrary root point fixed at the origin, 
The corresponding interpolated measure $d\nu$ can be described also very simply
in this picture. There is now a $\sigma_v$ field copy 
for every domain $v$ inside the big loop, 
a $w$ parameter for each dotted line,
and the covariance of two $\sigma_v$ and $\sigma_{v'}$ fields is the ordinary $\delta$ function 
covariance multiplied by a weakening parameter which is the infimum of the $w$ parameters of the 
dotted lines one has to \emph{cross} to go from $v$ to $v'$. The counterclockwise orientation
of the big loop corresponds to the $+iH$ convention.

In this new picture we see indeed many loops... but the golden rule is
not violated. In this new representation it simply translates into

\emph{``Thou shall see only planar (or genus-bounded) structures..."}

(Recall that genus-bounded graphs are not many and don't make perturbation
theory diverge.)

Let us prove now that the right hand side of formula 
(\ref{treeformul}) is convergent as series in $n$.

\begin{theorem}\label{goodtheor}
The series (\ref{treeformul}) is absolutely convergent for $\lambda$ small enough,
and the sum is bounded by $KM^{4j}$.
\end{theorem}

\noindent{\bf Proof}\ 
We shall use the first representation of Figure \ref{looptree}.
Consider a loop vertex $V_v$ of coordination $k_v$ in the tree.
Let us compute more explicitly the outcome of the $k_v$ derivatives 
$\prod_{i=1}^{k_v}\frac{\delta}{\delta \sigma(x_i)}$ acting on 
$$V= - \frac 12 Tr\log (1+iH)$$
which created this loop vertex.

Consider the operator 
\begin{equation}\label{resolventj}
C_{j}(\sigma) = D_j  \frac{1}{1+i H } D_j .
\end{equation}

Calling $x_1$ the root position for the loop vertex $V_v$, that is the unique
position from which a path goes to the root of $T$, the loop vertex factor $V_v$ 
after action of the derivatives is 
\begin{equation}\label{loopvertex}
[\prod_{i=1}^{k_v}\frac{\delta}{\delta \sigma(x_i)} ]    V_v =\frac 12   (-i \sqrt \lambda )^{k_v}   \sum_{\tau}\prod_{i=1}^{k_v}  C_j(\sigma, x_{\tau(i)} , x_{\tau(i+1)})  
\end{equation}
where the sum is over all permutations $\tau$ of $[2,...,k]$, completed by
$\tau(1) = \tau(k+1) =1$.

To bound the integrals over all positions except the root, we 
need only a very simple lemma:

\begin{lemma}\label{key}
There exists $K$ such that for any $x$ and any $v$  
\begin{equation}\label{pointbound}
\vert [ C_{j}(\sigma_v)]^{k_v} (x,x) \vert   \le   K^{k_v} M^{(4-2k_v)j}
\ \ \forall \sigma_v \; .
\end{equation}
\end{lemma}
Since $iH$ is anti-hermitian we have $\Vert (1+iH)^{-1}\Vert \le 1 $.
It is obvious from  (\ref{bound}) that $\Vert C_j \Vert \le K M^{-2j}$, hence 
$\Vert D_j \Vert \le K M^{-j}$.
We have  
\begin{equation}\label{normbound}
[ C_{j}(\sigma_v)]^{k_v} (x,x)  =
\int dy dz D_j (x,y) A (y,z)  D_j (z,x)  
=  <f, Af>
\end{equation}
for $f= D_j (x, .)$ and $A=(1+iH)^{-1} [ C_j  (1+iH)^{-1} ]^{k_v-1}$.
The norm of the operator $A$ is bounded by $ K^{k_v-1} M^{-2j(k_v -1)}$.
Since $\Vert f\Vert^2 \le K M^{2j}$, the result follows.
\qed

To bound the $dx_\ell$ integrals we start from the leaves
and insert the bound (\ref{pointbound}), which also means that the multiplication
operator $C_{j}(\sigma_v)]^{k_v} (x,x)$ (diagonal in $x$ space) has a norm
bounded by  $K^{k_v} M^{(4-2k_v)j}$ uniformly in $\sigma$. 
We then progress towards the root. 
By induction, multiplying norms, adding the 
$\frac 12   (-i \sqrt \lambda )^{k_v}$ factors from (\ref{loopvertex}) and taking into account
the factorials from the sum over the permutations  $\tau$ in  (\ref{loopvertex})
gives exactly 
\begin{equation}
\label{treeboundx} 
\prod_v  \frac 12 (k_v-1)! \lambda^{k_v/2} K ^{k_v}M^{4j-2jk_v}.
\end{equation}

For a tree on $n$ loop vertices
$\sum_v k_v = 2(n-1) $ hence $\sum_v  (4-2k_v) = 4n -4(n-1) =4 $ so that collecting all dimensional 
factors we get a $M^{4j}$ global $n$ independent factor as should be the case
for vacuum graphs in the $\phi^4$ theory in a single RG slice.

We can now integrate the previous bound over the complicated measure $d\nu_T$
and over the $\{w_\ell\}$ parameters. 
But since our bound is independent of ${\sigma^v}$, since the measure $d\nu(\sigma)$ 
is normalized, and since each $w_\ell$ runs from 0 to 1, this does not change the result.

Finally by Cayley's theorem the sum over trees costs $\frac {n!}{ \prod_v (k_v -1)!}$.
The $n!$ cancels with the $1/n!$ of (\ref{treeformul}) and the $1/(k_v-1) !$ exactly cancel the 
one in (\ref{treeboundx}) . It remains a geometric series bounded by 
$\frac 12 M^{4j} (\lambda K)^{n-1}$ hence convergent for
small $\lambda$, and the sum is bounded by $K. M^{4j}$.
\qed

\section{Uniform Borel summability}

Rotating to complex $\lambda$ and Taylor expanding out a fixed number of $\phi^4$ vertices proves 
Borel summability in $\lambda$ \emph{uniformly in} $j$. 

\medskip
\noindent{\bf Definition}
\medskip
{\it
A family $f_j$ of functions is called {\bf Borel summable} in $\lambda$ uniformly in $j$  if 

\begin{itemize}
\item
Each $f_j$ is analytic in a disk 
$D_R = \{ \lambda \vert {\rm Re}\, \lambda^{-1} > 1/R\}$;

\item Each $f_j$ admits an asymptotic power series $\sum_k a_{j,k} \lambda^k  $
(its Taylor series at the origin) hence:
\begin{equation}  f_j(\lambda) = \sum_{k=0}^{r-1} a_{j,k} \lambda^k + R_{j,r} (\lambda)  
\end{equation} 
such that the bound
\begin{equation} \label{taylorremainder}  \vert R_{r,j} (\lambda) \vert \le A_j \rho^r r! \vert \lambda \vert^r 
\end{equation} 
holds uniformly in $r$ and $\lambda \in D_R$, for some constant $\rho \ge 0$ independent
of $j$ and constants $A_j \ge 0$ which may depend on $j$.

\end{itemize}
}

Then every $f_j$ is Borel summable \cite{Sok}, i.e. the power series 
$\sum_k a_{j,k} \frac{t^k }{ k!}$ converges for $\vert t \vert < \frac{1 }{\rho}$, it 
defines a function $B_j(t)$ which has an analytic continuation in the $j$ independent strip 
$S_{\rho} = \{t \vert {\rm \ dist \ } (t, {{\mathbb R}}^+) < \frac{1}{ \rho}\}$.
Each such function satisfies the bound
\begin{equation} \vert B_j(t)  \vert \le { \rm B_j} e^{\frac{t }{R}} \quad {\rm for \ } 
t \in { {\mathbb R}}^+  
\end{equation}
for some constants $B_j \ge 0$ which may depend on $j$.
Finally each $f_j$ is represented by the following absolutely convergent integral:
\begin{equation}  f_j(\lambda) = \frac{1 }{ \lambda} \int_{0}^{\infty} e^{-{\frac{t} {\lambda}} } B_j(t) dt \quad\quad
 \quad {\rm for \ } \lambda \in C_R .
\end{equation}

\begin{theorem}
The series for the pressure is uniformly Borel summable with respect
to the slice index.
\end{theorem}

\noindent{\bf Proof}
It is easy to obtain uniform analyticity for ${\rm Re}\, \lambda >0$ and $\vert \lambda\vert $
small enough, a region which obviously contains a disk $D_R$.
Indeed all one has to do is to reproduce the previous argument but 
adding that for $H$ Hermitian, the operator
$(1+i e^{i \theta} H)^{-1}$ is bounded by $\sqrt 2$ for $\vert \theta \vert \le \pi /4$.
Indeed if $\pi/4 \le {\rm Arg} z \le 3\pi/4 $, we have $\vert (1+i z )^{-1}\vert \le \sqrt 2$.

Then the uniform bounds (\ref{taylorremainder}) follow from 
expanding the product of resolvents in (\ref{loopvertex}) up to order $r-2(n-1)$
in $\lambda$ by an explicit Taylor formula with integral remainder followed 
by explicit Wick contractions. The sum over the contractions leads to the 
$\rho^r r!$ factor in (\ref{taylorremainder}).
\qed

\section{Connected functions and their decay}

\label{iterresol}

To obtain the connected functions with
external legs we need to add resolvents to the initial loop vertices.  A resolvent is
an operator $C_{j}(\sigma_r, x, y ) $. The connected functions $S^c(x_1, ..., x_{2p}) $
are obtained from the normalized functions by the standard procedure. We 
have the analog of formula (\ref{treeformul}) for these connected functions:
\begin{theorem}

\begin{eqnarray}\label{treeformulext}
S^{c}(x_1, ..., x_{2p}) 
&=& \sum_{\pi} \sum_{n=1}^{\infty}\frac{1}{n!} \sum_T \bigg\{ \prod_{\ell\in T}   
\big[ \int_0^1 dw_\ell \int d^4 x_\ell d^4 y_\ell \big]\bigg\}  \nonumber \\
&&\hskip-3.5cm\int  d\nu_T (\{\sigma_v\}, \{\sigma_r\}, \{ w \}) 
\bigg\{ \prod_{\ell\in T} \big[ \delta (x_\ell - y_\ell) 
 \frac{\delta}{\delta \sigma_{v(\ell)}(x_\ell)}\frac{\delta}{\delta \sigma_{v'(\ell)}(y_\ell)} 
 \big] \bigg\} \nonumber \\
 && \prod_v V_v \prod_{r=1}^{p}   C_{j}(\sigma_{r}, x_{\pi(r,1)}, x_{\pi(r,2)})\; ,
\end{eqnarray}
where 
\begin{itemize}
\item the sum over $\pi $ runs over the pairings of the $2p$ external variables
into pairs $(x_{\pi(r,1)}, x_{\pi(r,2)})$, $r=1,...,p$, 

\item each line $\ell$ of the tree joins two different loop vertices or resolvents 
$V_{v(\ell)}$ and $V_{v'(\ell)}$ 
at point $x_{\ell}$ and $y_{\ell}$, which are identified through the function
$\delta (x_\ell - y_\ell) $ because the covariance of $\sigma$ is ultralocal,

\item the sum is over trees joining the $n+p$ loop vertices and resolvents, which have therefore
$n+p-1$ lines,

\item the measure $d\nu_T (\{\sigma_v\}, \{\sigma_r\}, \{ w \})  $ over the 
$\{\sigma\}$ fields has covariance 
\ \ $<\sigma_\alpha,\sigma_{\alpha'}>=
\delta (x-y) w^T (\alpha, \alpha', \{ w\})$ where $w^T (\alpha, \alpha', \{ w\})$ is 1 if $\alpha=\alpha'$ (where $\alpha, \alpha'\in \{v\}, \{r\}$),
and the infimum of the $w_\ell$ for $\ell$ running over the unique path from $\alpha$ to $\alpha'$ in $T$
if $\alpha\ne \alpha'$. This measure is well-defined because the matrix $w^T$ is positive.

\end{itemize}
\end{theorem}

Now we want to prove not only convergence of this expansion
but also scaled tree decay between external arguments:

\begin{theorem}
The series (\ref{treeformulext}) is absolutely convergent for $\lambda$ small enough, 
its sum is uniformly Borel summable in $\lambda$ and we have:
\begin{equation}\label{decaybound}
\vert  S^{c}(z_1, ..., z_{2p}) \vert \le (2p)!  K^p \vert \lambda \vert^{p-1}  
M^{2p j} e^{-cM^j d(z_1,...,z_{2p})}
\end{equation}
where $d(z_1,...,z_{2p})$ is the length of the shortest tree which connects all the
points $z_1, ..., z_p$.
\end{theorem}

The proof of convergence (and of uniform Borel summability) is similar to the one for the pressure.

The tree decay (\ref{decaybound}) is well known and standard to establish 
through the traditional cluster and Mayer expansion. It is due 
to the existence of a tree of $C_j$ propagators between external points in any connected function. 
In the present expansion, this tree is hidden in the resolvents and loop vertices,
so that an expansion on these resolvents (and loop vertices) is necessary in one form 
or another to prove (\ref{decaybound}). It does not seem to 
follow from bounds on operator norms only: the integral over the 
$\sigma$ field has to be bounded more carefully.

The standard procedure to keep resolvent expansions convergent 
is a so-called large/small field expansion on $\sigma$. In the region where
$\sigma $ is small the resolvent expansion converges. In the large field region
there are small probabilistic factors coming from the $d \nu_T$ measure. This is 
further sketched in
subsection \ref{largefield}.

However the large/small field expansion again requires a discretization of space into 
a lattice: a battery of large/small field tests is performed, on the average of the field
$\sigma $ over each cube of the lattice.
We prefer to provide a  new and different proof of (\ref{decaybound}). It relies 
on a single resolvent step followed by integration by parts, to establish
a Fredholm inequality on the modulus square of the $2p$ point function.
From this Fredholm inequality the desired decay follows easily.
The rest of this section is devoted to the proof of (\ref{decaybound})
in the simplest case $p=1$.  The most general case is sketched in subsection \ref{higher}.

\begin{figure}[!htb]
\centering
\includegraphics[scale=0.5]{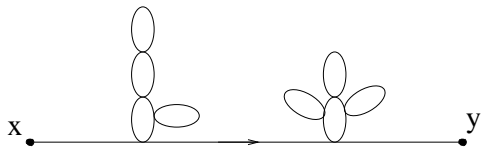}
\caption{Three resolvents with two branching subtrees}
\label{treecactus}
\end{figure}

The two point function $S^c$ is simply called $S(x,y)$ from now on,
and for $p=1$ (\ref{decaybound}) reduces to
\begin{equation}\label{decayboundbis}
\vert  S(x,y) \vert \le  K  M^{2 j} e^{-cM^j \vert x-y \vert}.
\end{equation}
We work with $n$, $T$ and $\{w\}$ fixed in (\ref{treeformulext}).
We use the resolvent as root for $T$, from which grow $q$ subtrees $T_1, ... , T_q$.
In more pictorial terms,  (\ref{treeformulext})  
represents a chain of resolvents from $x$ to $y$ separated by insertions of $q$ subtrees. Figure 
\ref{treecactus} is therefore the analog of Figure \ref{looptree}
in this context\footnote{A similar figure is a starting point for the 1PI expansion of the self-energy
in \cite{DR1,Hub}.}.

A representation similar to the big loop of Figure \ref{bigloop}
pictures the decorated resolvent as a half-circle
going from $x$ to $y$, together with a set of planar dotted lines for the vertices. 
The $+i$ convention again corresponds to a particular orientation. For reason
which should become clear below, we picture the planar 
dotted lines all on the same side of the $x$-$y$ line, hence 
\emph{inside the half-disk}.

\begin{figure}[!htb]
\centering
\includegraphics[scale=0.5]{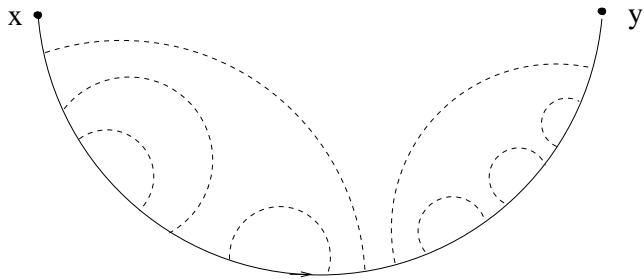}
\caption{The half-circle representation of Figure \ref{treecactus}}
\label{resolvent}
\end{figure}

To each such drawing, or graph $G$, there is an associated Gaussian measure $d\nu_G$
which is the one from which the drawing came as a tree. Hence it 
has a field copy associated  to each planar region of the picture, 
a weakening parameter $w$ associated to each dotted line,
and the covariance between the $\sigma$ fields of different regions 
is given by the infimum over the parameters of the dotted lines
that one has to cross to join these two regions.

There is also for each such $G$ an \emph{amplitude}. Let us write simply
$\int d\nu_G$ for the normalized integral 
$\int_0^1 \prod_{\ell \in G} dw_\ell \int  d\nu_G (\{\sigma\}, \{ w \})  $. If the graph has $n$
dotted lines hence $2n+1$ resolvents from $x$ to $y$, its amplitude is
\begin{eqnarray}\label{amplitude}
A_G (x,y) &= & \lambda^n
\int  d\nu_G 
 \int  \big[ \prod_{\ell \in G} d^4 x_\ell \big] \prod_{i=1}^{2n+1}  
C_{j}(\sigma_{i}, x_{i-1}, x_{i})
\end{eqnarray}
where the product over $\ell$ runs over the dotted lines and the product
over $i$ runs over the resolvents along the half-circle, with $x_0=x$ and
$x_{2n+1}=y$. $\sigma_i$ is the field copy of the region just before point
$x_i$ and the $2n$ positions $x_1, ..., x_{2n}$ are equal in pairs to the 
$n$ corresponding $x_\ell$'s according to the pairings 
of the dotted lines.

We shall prove
\begin{lemma}\label{firstlemma}
There exists some constant $K$ such that for $\lambda$
small enough 
\begin{equation}\label{twopointbound}
\sup_{G, n(G) = n} \vert A_{G}(x,y) \vert  \le  (\vert \lambda\vert K )^{n/2}  
M^{2j} e^{-cM^j \vert x - y \vert }.
\end{equation}
\end{lemma}
From this Lemma (\ref{decayboundbis}) obviously follows.  Indeed the remaining sum over Cayley trees
costs at most $K^n n!$, which is compensated by the $\frac {1}{n!}$ in
(\ref{treeformulext}). In the language of planar graphs the planar dotted lines
cost only $K^n$. Hence the sum over $n$ converges for $\lambda$ small enough
because of the $\vert \lambda\vert^{n/2} $ factor in (\ref{twopointbound}). 
Remark that this factor $\vert \lambda\vert ^{n/2}  $ is not optimal;
$\vert \lambda\vert^{n}  $ is expected; but it is convenient to use
half of the coupling constants for auxiliary sums below.

We apply a Schwarz inequality to $\vert A_{G}(x,y) \vert^2 $, relatively
to the normalized measure $d\nu_G$:
\begin{eqnarray}   \label{squareamp}
\vert A_{G}(x,y) \vert^2  & \le & A_{G  \cup \bar G} (x,y), 
 \\ \nonumber  
A_{G \cup \bar G} (x,y) &=& \int  d\nu_G    
 \int \big[\prod_{\ell \in G} d^4 x_\ell  d^4 \bar x_\ell \big] 
\\  &&\prod_{i=1}^{2n+1}  
C_{j}(\sigma_{i}, x_{i-1}, x_{i})\bar C_{j}(\sigma_{i}, \bar x_{i-1},\bar x_{i})
\label{posamp}
\end{eqnarray}
with hopefully straightforward notations.

\begin{figure}[!htb]
\centering
\includegraphics[scale=0.5]{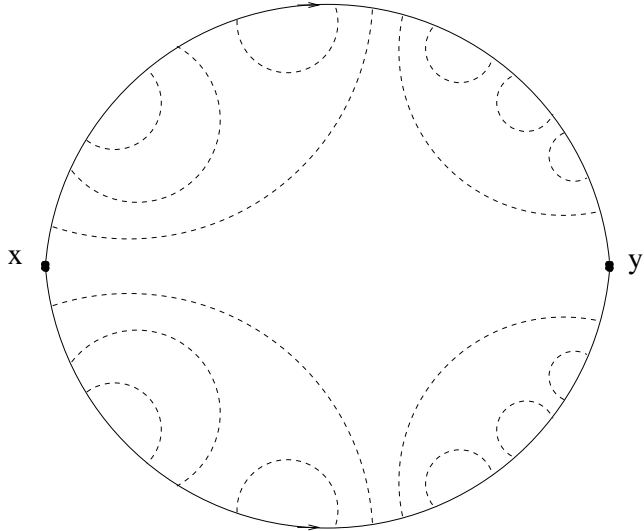}
\caption{The mirror graph  $G\cup \bar G$ for the graph $G$ of Figure \ref{resolvent}}
\label{mirror}
\end{figure}

The quantity on the right hand side is now pointwise positive for any $\sigma$.
It can be considered as the amplitude $A_{G \cup \bar G} (x,y)$
associated to a \emph{mirror graph}  $G \cup \bar G$.
Such a mirror graph is represented by a full disk, 
with $x$ and $y$ diametrally opposite, and no dotted line crossing 
the corresponding diameter. The upper half-circle represents the complex 
conjugate of the lower part. Hence the upper half-disk is exactly the mirror 
of the lower half-disk, with orientation reversed, see Figure \ref{mirror}.

The Gaussian measure associated to such a mirror graph 
remains that of $G$, hence it has a single weakening $w$ parameter for each dotted line and its mirror line, and it has a single
copy of a  $\sigma$ field for each \emph{pair} made of a region of the disk \emph{and its mirror
region}. Let's call such a pair a ``mirror region". 
The covariance between two fields belonging to two mirror regions
is again the infimum of the $w$ parameters crossed from one region to the other,
but e.g. staying entirely in the lower half-disk (or the upper half-disk).

\begin{figure}[!htb]
\centering
\includegraphics[scale=0.5]{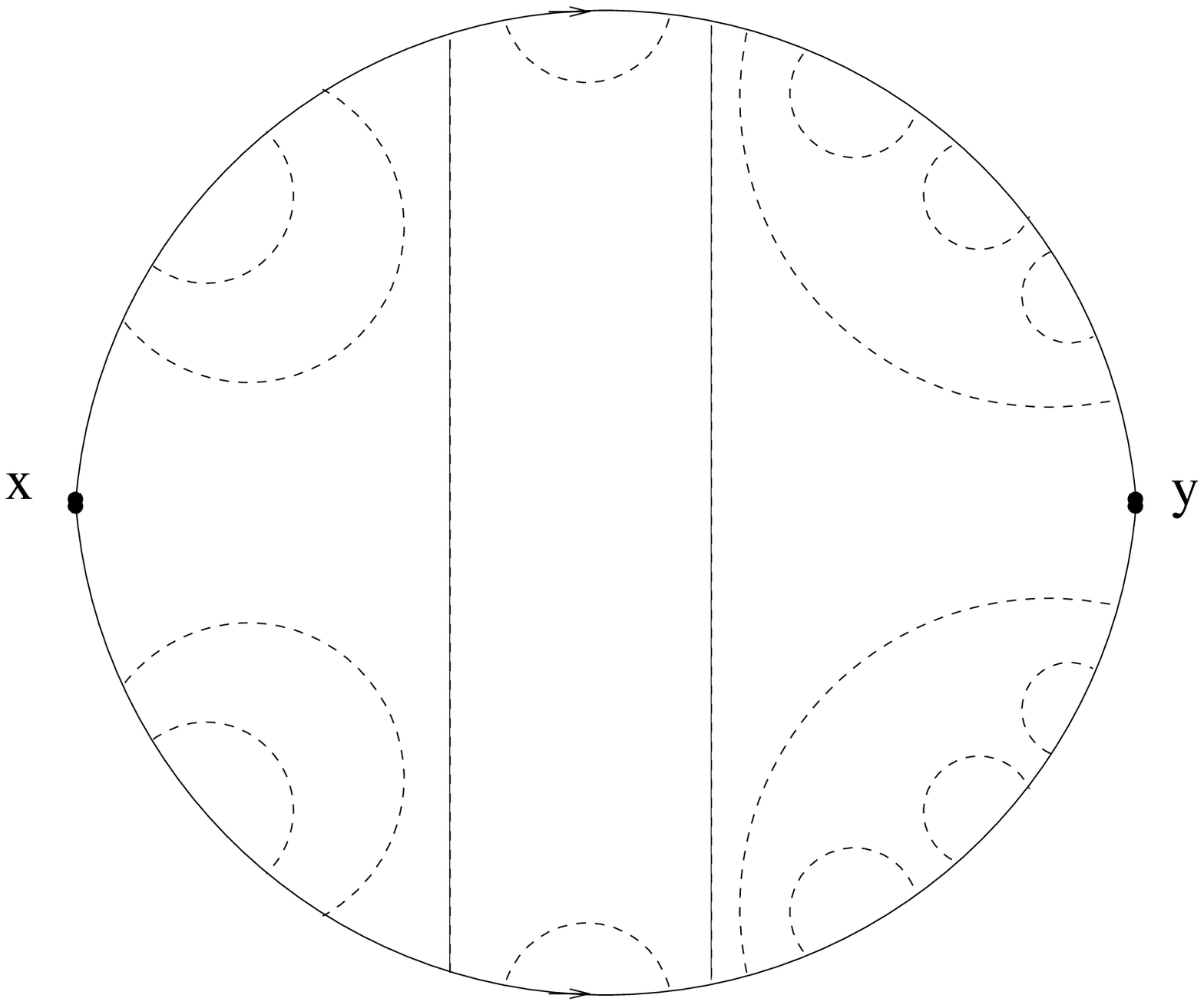}
\caption{The generalized mirror graphs}
\label{genmirror}
\end{figure}

We shall now perform a single resolvent expansion step and integration by parts,
together with a bound which reproduces an amplitude similar to $A_{G \cup \bar G}$.
The problem is that the category of mirror graphs is not exactly stable in this operation;
this bound generates other graphs with ``vertical" dotted lines between the lower and upper half of the circle. To prove our bound inductively
we need therefore to generalize slightly
the class of \emph{mirror graphs} and their associated Gaussian measures
to a larger category of graphs $G\cup \bar G \cup V$, called  \emph{generalized mirror graphs} 
or GM graphs and pictured in Figure \ref{genmirror}. They are identical
to mirror graphs except that they can have in addition a certain set $V$ of ``vertical"
dotted lines between the lower and upper half of the circle, again without any crossing.

There is a corresponding measure $d\nu_{G,V}$
with similar rules; there is a single $w$ parameter for each pair of dotted line and its mirror,
in particular there is a $w$ parameter for each vertical line, Again
the covariance between two fields belonging to two 
mirror regions is the infimum of the $w$ parameters crossed from one mirror
region to the over,  \emph{staying entirely in e.g. the lower half-disk}.
The upper half-part is still the complex conjugate of the lower half-part. 
The order of a GM graph is again the total number $L= 2n +\vert V \vert $ of dotted lines
and its amplitude is given by a pointwise positive integral similar to (\ref{posamp}): 

\begin{eqnarray}\label{amplitudebig}
A_{G\cup \bar G \cup V} (x,y) &=& \lambda^L  \int d\nu_{G \cup V}
\int \big[ \prod_{\ell\in G} 
 d^4 x_\ell d^4 \bar x_\ell  \big] \big[\prod_{\ell\in V} dy_\ell \big] \nonumber
\\
&&  \prod_{i=1}^{2n+\vert V \vert+1}  
C_{j}(\sigma_{i}, z_{i-1}, z_{i}) 
\bar C_{j}(\sigma_{i}, \bar z_{i-1}, \bar z_{i}) ,
\end{eqnarray}
where the $z$'s and $\bar z$'s  
are either $x_\ell$'s, $\bar x_\ell$'s or $y_\ell$'s according to the graph.

Defining the integrand $I_{G\cup \bar G \cup V}(x,y)$ of a GM graph so that
$A_{G\cup \bar G \cup V} (x,y) =\int d\nu_{G \cup V}  I_{G\cup \bar G \cup V}(x,y) $,
we have:

\begin{lemma}
For any GM graph we have, uniformly in $\sigma$, $x$ and $y$:
\begin{eqnarray}\label{gmgnorm}
I_{G\cup \bar G \cup V} (x,y)  \le (K\vert  \lambda \vert )^L M^{4j} .
\end{eqnarray}
\end{lemma}
Inded the quantity $I_{G\cup \bar G \cup V} (x,y) $ is exactly the same than a pressure
graph but with two fixed points and some propagators replaced by complex conjugates, 
hence the proof through the norm estimates of Lemma \ref{key} 
is almost identical to the one of Theorem \ref{goodtheor}.

We now write the resolvent step
which results in an integral Fredholm inequality for the supremum
of the amplitudes of any generalized mirror graph.

Let us define the quantity 
\begin{equation}\label{defgamma}
\Gamma_L (x,y) = \sup_{GM \ {\rm graphs} \ G,V \ | \ L(G) = L} \vert  \lambda\vert^{-L/2}
A_{G\cup \bar G \cup V} (x,y) .
\end{equation}
We shall prove by induction on $L$:
\begin{lemma}
There exists some constant $K$ such that for $\lambda$
small enough 
\begin{eqnarray}\label{wantedfred}
\Gamma_L (x,y ) &\le& 
K  M^{4j} \bigg( e^{-cM^{j} \vert x-y \vert} +  \vert  \lambda \vert^{1/2} 
\int dz  e^{-cM^{j} \vert x-z \vert}  \Gamma_L (z,y ) \bigg).
\end{eqnarray}
\label{secondlemma}
\end{lemma}
From that lemma indeed obviously follows 
\begin{lemma}\label{thirdlemma}
There exists some constant $K$ such that for $\lambda$
small enough 
\begin{eqnarray}\label{wanteddec}
\Gamma_L (x,y ) &\le& K  M^{4j} e^{-cM^{j} \vert x-y \vert}.
\end{eqnarray}
\end{lemma}
Indeed iterating the integral Fredholm equation 
(\ref{wantedfred}) leads obviously to (\ref{wanteddec}).

Taking (\ref{amplitudebig}) and (\ref{defgamma}) into account to reinstall
the $\lambda^{L/2}$ factor, considering
the equation $L=2n +V$ and taking a square root because of (\ref{squareamp}), 
Lemma \ref{firstlemma} 
is then nothing but Lemma \ref{thirdlemma} for the particular case $V=0$.

\medskip
The rest of this section is therefore devoted to the proof of Lemma 
\ref{secondlemma}, by a simple induction on $L$.
\medskip

If $L =0$, $\Gamma_0 (x,y) = \int d\nu C_j (\sigma, x,y,) \bar C_j (\sigma, x,y,)$.
Expanding the $C_j (\sigma, x,y)$ propagator, we get 
\begin{eqnarray}
\Gamma_0 (x,y) = \int d\nu \big[ C_j(x, y) -  i \sqrt{\lambda} \int  dz C_j(x, z) \sigma (z)  C_j(\sigma, z, y)\big] \bar C_j (\sigma, x,y).
\end{eqnarray}
For the first term $\vert \int d\nu  C_j(x, y)  \bar C_j (\sigma, x,y) \vert  $, 
we simply use bounds (\ref{bound}) and  
(\ref{gmgnorm}) in the case $L=0$. For the second term
we Wick contract the $\sigma$ field (i.e. integrate by parts over $\sigma$). 
There are two subcases: the 
Wick contraction $\frac {\delta}{\delta \sigma}$ hits either $ C_j(\sigma, z, y)$ or $\bar C_j(\sigma, x, y)$.
We then apply the inequality 
\begin{eqnarray} \label{ineq}
\vert ABC \vert \le \frac {A}{2}( M^{2j} \vert B \vert^2 + M^{-2j} \vert C \vert^2  ),
\end{eqnarray}
which is valid for any positive $A$. 
In the first subcase we take $A=\int dz C_j(x, z) $, $B=C_j(\sigma, z, y) $ 
and $C=C_j(\sigma, z, z)\bar C_j(\sigma, x, y)$,
hence write
\begin{eqnarray}
&&\hskip -1cm \vert \int dz C_j(x, z) C_j(\sigma, z, z) C_j(\sigma, z, y) \bar C_j(\sigma, x, y) \vert
\le \nonumber  \\
 &&\int dz \frac {C_j(x, z)}{2} \big[ M^{2j}\vert  C_j(\sigma, z, y) \vert^2  +
 M^{-2j} \vert  C_j(\sigma, z, z)\bar C_j(\sigma, x, y)\vert^2  \big] 
\end{eqnarray}
and in the second subcase we write similarly
\begin{eqnarray}
&&\hskip -1cm 
\vert\int dz  C_j(x, z) C_j(\sigma, z, y) \bar C_j(\sigma, x, z) \bar C_j(\sigma, z, y)\vert
\le   \nonumber   \\
&&\int dz \frac { C_j(x, z)}{2} \big[ M^{2j}\vert C_j(\sigma, z, y) \vert^2  
+ M^{-2j}  \vert  \bar C_j(\sigma, x, z)\bar C_j(\sigma, z, y) \vert^2 \big] .
\end{eqnarray}
Using the uniform bound  (\ref{gmgnorm}) on the ``trapped loop"  $\vert C_j(\sigma, z, z)\vert^2$
or  $\bar C_j(\sigma, x, z)\vert^2$  in the $C$ term we obtain
\begin{eqnarray}
\Gamma_0 (x,y ) &\le&  K M^{4j} e^{-cM^{j} \vert x-y \vert} +  \vert \lambda\vert K  \bigg(
\Gamma_0 (x,y ) 
\nonumber \\
&&+  M^{4j} \int  dz  e^{-cM^{j} \vert x-z \vert} \Gamma_0 (z,y) \bigg)
\end{eqnarray}
so that (\ref{wantedfred}) hence Lemmas \ref{secondlemma} and \ref{thirdlemma} hold 
for $L=0$.

We now assume that (\ref{wantedfred}), hence also (\ref{wanteddec}), is true up to order 
$L$ and we want to prove (\ref{wantedfred}) at order $L+1$.
Consider a GM graph of order $L+1$. If $V \ge 1$ 
we can decompose it as a convolution of smaller GM graphs:
\begin{eqnarray}
A_{G\cup \bar G \cup V} (x,y)
=\lambda \int dy_1 A_{G_1\cup \bar G_1} (x,y_1) A_{G_2\cup \bar G_2 \cup V_2} (y_1,y) 
\end{eqnarray}
with total orders $L_1$ for $G_1$ and $L_2$ for $G_2, V_2 = V-\{1\}$ strictly smaller
than $L+1$. Applying the induction hypothesis (\ref{wanteddec}) to these smaller
GM graphs we get directly that 
\begin{eqnarray}
\sup_{G,V | L(G\cup \bar G \cup V) = L+1, V >0} 
\vert  \lambda\vert^{-(L+1)/2} A_{G\cup \bar G \cup V} (x,y)
\le  K  M^{4j} e^{-cM^{j} \vert x-y \vert} .
\end{eqnarray}

Hence we have now only to prove (\ref{wantedfred}) for mirror graphs with $V=\emptyset$. 
Consider now such a mirror graph $G$.
Because of the $\vert \lambda \vert^{-L/2} $ in (\ref{defgamma}),
we should remember that we have only a remaining factor $\vert \lambda \vert^{L/2} $
to use for our bounds on $\Gamma_L$.

Starting at $x$ we simply expand the first resolvent propagator $C_j( \sigma , x, x_1 )$ as 
$ C_j(x, x_1) - \int  dz C_j(x, z) i \sqrt{\lambda}\sigma (z)  C_j(\sigma , z,x_1 )$.

For the first term we call $x_{i_1}$ the point to which $x_1$ is linked by a dotted line and
apply a Schwarz inequality of the (\ref{ineq}) type, with:
\begin{eqnarray}
A&=& \int dx_1   C_j(x, x_1)  , \\
B&=&  \int\prod_{i_1+1 \le i \le 2n} dx_i 
\prod_{i_1+1 \le i \le 2n+1 } C_j(\sigma, x_{i-1}, x_{i} ), 
\nonumber \\  \nonumber 
C&=&    \int \prod_{2\le i \le i_1 -1}  dx_i  \prod_{2\le i \le i_1}  C_j(\sigma, x_{i-1}, x_{i})    
\prod_{i=1}^{2n} d\bar x_i \prod_{1\le i \le 2n+1 } \bar C_j(\sigma, \bar x_{i-1}, \bar x_{i}).
\end{eqnarray}
It leads, using again 
the norm bounds of type (\ref{gmgnorm}) on the ``trapped loop" in the first part of $C$, to a bound
\begin{eqnarray}\label{firsttermbound}
\vert \lambda\vert^{1/2} K  \bigg(\Gamma_L (x,y ) 
+ M^{4j} \int  dx_1  e^{-cM^{j} \vert x-x_1 \vert} \Gamma_r (x_1,y) \bigg)
\end{eqnarray}
for some $r < L$. Applying the induction hypothesis concludes to the bound (\ref{wantedfred}). 

Finally for the second term we Wick contract again the $\sigma$ field. There are again two subcases: the 
Wick contraction $\frac {\delta}{\delta \sigma}$ hits either a $ C_j $ or a $\bar C_j $.
Let us call $i$ the number of half-lines, either on the upper or on the lower circles,
which are inside the Wick contraction, and
$x_{i_1}$, ... $x_{i_k}$ or $\bar x_{i_1}$, ... $\bar x_{i_k}$
the positions of the dotted lines \emph{crossed´} 
by the Wick contraction. 
 
We have now two additional difficulties compared to the $L=0$ case: 

\begin{itemize}

\item we have to sum over where the Wick contraction hits, hence sum over $i$
(because the Wick contraction creates a loop, hence potentially dangerous combinatoric). 
The solution is that the norm bound on the ``trapped loop" in the $C$ term
 of (\ref{ineq}) erases more and more coupling constants as the loop gets longer: this
easily pays for choosing the Wick contraction.

\item the dotted lines \emph{crossed} by the Wick contraction 
should be kept in the $A$ term in inequality (\ref{ineq}). In other words 
they become vertical lines at the next step, even if no vertical line was present 
in the initial graph. This is why we had to extend our induction to the category
of GM graphs. This extension is what solves this difficulty. 

\end{itemize}

\begin{figure}[!htb]
\centering
\includegraphics[scale=0.5]{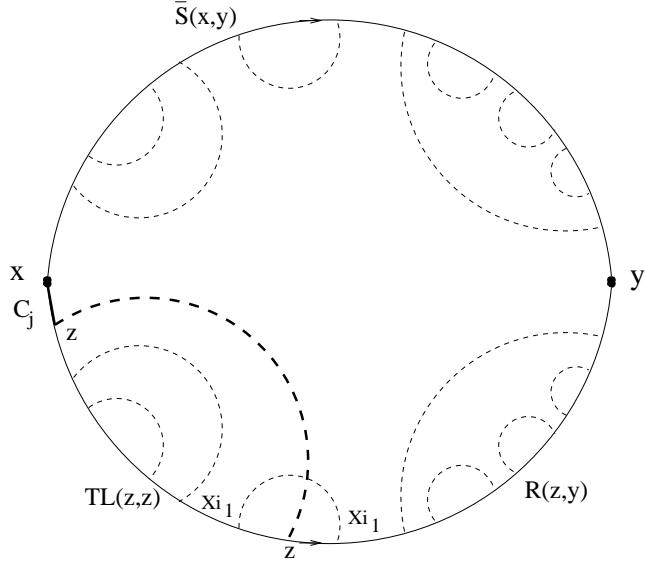}
\caption{The Wick contraction}
\label{mirrorwick}
\end{figure}

We decompose the amplitude of the graph  in the first subcase of Figure \ref{mirrorwick}
as
\begin{eqnarray}
\sum_i \int dz dx_{i_1}, ... dx_{i_k} 
C_j(x, z) TL_{x_{i_1}, ...x_{i_k}} (z, z)   R_{x_{i_1}, ...x_{i_k}}(z, y) 
\bar S (x, y)
\end{eqnarray}
with hopefully straightforward notations, and we apply the Schwarz inequality (\ref{ineq}), with:
\begin{eqnarray}
A&=&  \vert \lambda \vert^{i/8}
\sum_i \int dz dx_{i_1}, ... dx_{i_k} \int   C_j(x, z) ,   \nonumber \\
B&=&  R_{x_{i_1}, ...x_{i_k}}(z, y) ,   \nonumber \\
C&=&\vert \lambda \vert^{-i/8} TL_{x_{i_1}, ...x_{i_k}} (z, z)   \bar S (x, y) .
\label{finalABC}
\end{eqnarray}

Now the first remark is that $i \vert \lambda \vert^{i/8}$ is bounded by $K$ for
small $\lambda$ so we need only to find a uniform bound at fixed $i$.

The $A\vert B\vert ^2$ is a convolution of an explicit propagator bounded
by (\ref{bound}) with a new GM graph (with vertical lines
which are the crossed lines at $x_{i_1}, ...x_{i_k}$)
either identical to $G$ or shorter. If it is shorter 
we apply the induction hypothesis. If it is not shorter we obtain a convolution equation
term like in the right hand side of (\ref{wantedfred}).

The $A\vert C\vert ^2$ contains a trapped loop $TL$ with $i$ vertices. Each half-vertex
of the trapped loop has only $\vert \lambda \vert^{1/8}$
because of the $\vert \lambda \vert^{-i/8}$ factor in (\ref{finalABC}). 
The trapped loop is again of the GM nature 
with vertical lines which are the crossed lines at $x_{i_1}, ...x_{i_k}$.
But we can still apply the bound
(\ref{gmgnorm}) to this trapped loop. Therefore the bound on the sum of the
$A\vert B\vert ^2$ and  $A\vert C\vert ^2$  is again of the type
(\ref{firsttermbound}).
 
Finally the second subcase, where the Wick contraction 
$\frac {\delta}{\delta \sigma}$ hits a $\bar C_j $, is exactly similar,
except that the ``almost trapped loop" is now something of the
type $\bar TL(x,z)$ rather than $TL(z,z)$. But the bound (\ref{gmgnorm}) 
also covers this case, so that everything goes through.

Collecting the bounds (\ref{firsttermbound}) in every case 
completes the proof of Lemmas 
\ref{secondlemma} and \ref{thirdlemma} for $\Gamma_{L+1}$.
This concludes the proof of Lemmas \ref{secondlemma} and \ref{thirdlemma} for all $L$.

\section{Further topics}

\subsection{Higher functions}\label{higher}

\begin{figure}[!htb]
\centering
\includegraphics[scale=0.5]{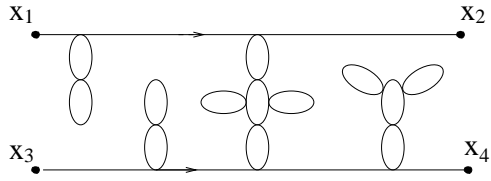}
\caption{A connected 4 point function}
\label{4cactus}
\end{figure}

\begin{figure}[!htb]
\centering
\includegraphics[scale=0.5]{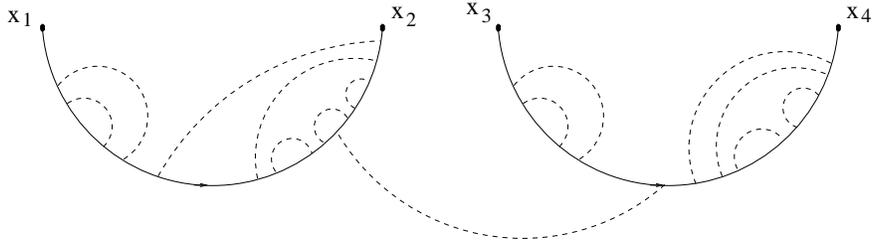}
\caption{The ``half-disk" representation of that connected 4 point function}
\label{4point}
\end{figure}

The analysis of the $2p$ point functions is similar to that of the previous section. 
The general $2p$ point function $S^c (x_1, ..., x_{2p})$ 
defined by (\ref{treeformulext}) 
contains $p$ resolvents of the $C_j (\sigma)$ type
and a certain number of loop vertices joining or decorating them.
Turning around the tree we can still identify the drawing as 
a set of decorated resolvents joined by local vertices
or dotted lines as in Figures  \ref{4cactus} and  \ref{4point}, which are the analogs of
Figures \ref{treecactus} and  \ref{resolvent}. This is because any chain of loop vertices
joining resolvents can be ``absorbed" into decorations of one of these resolvents. 

\begin{figure}[!htb]
\centering
\includegraphics[scale=0.5]{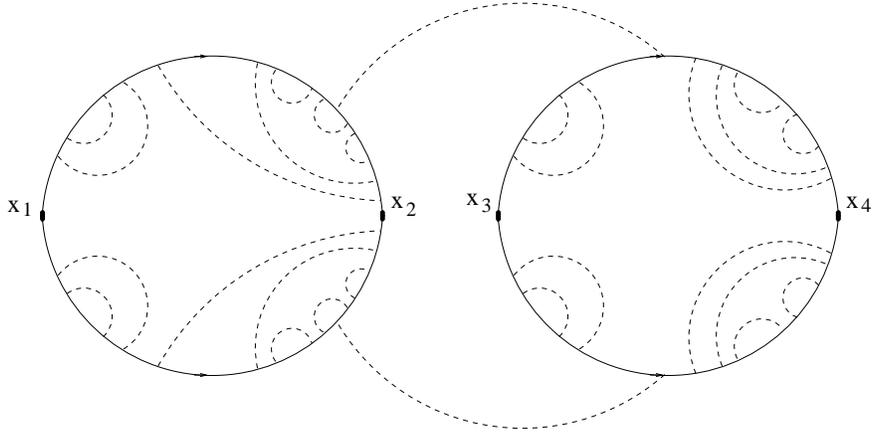}
\caption{The mirror representation of the same connected 4 point function}
\label{4pointmirror}
\end{figure}

The factor $2p!$ in (\ref{decaybound})
can be understood as a first factor $2p!!$ to choose the pairing of the points in $p$
resolvents and an other $p!$ for the choice of the tree of connecting loop vertices
between them. We can again bound each term of the initial expansion by a ``mirror" term
pointwise positive in $\sigma$ with $p$ disks as shown in Figure \ref{4pointmirror}.

A Lemma similar to Lemma \ref{firstlemma}
is again proved by a bound on generalized 
mirror graphs such as those of Figure \ref{4pointmirror} but
with additional vertical lines inside the $p$ disks. This bound
is proved inductively by a single resolvent step followed by a Fredholm bound 
similar to Lemmas \ref{secondlemma} and \ref{thirdlemma}.
Verifications are left to the reader.

\subsection{Large/small Field Expansion}\label{largefield}

To prove the tree decay of the $2p$-point connected functions as external
arguments are pulled apart, it is possible to replace the Fredholm inequality 
of the previous section by a so-called \emph{large/small field expansion}.
It still relies on a resolvent expansion, but integration by parts is
replaced by a probabilistic analysis over $\sigma$. We recall only the main idea,
as this expansion is explained in detail in \cite{AR3, KMR} but also
in a very large number of other earlier publications.

A lattice ${\cal D}$ of cubes of side $M^{-j}$ is introduced and the expansion is
\begin{eqnarray}
 1 = \prod_{\Delta \in {\cal D}} \bigg\{ \chi(  \int_{\Delta} M^{4j} \vert \lambda \vert^{\epsilon}\sigma^2 (x) dx ) + 
 [1-  \chi(  \int_{\Delta} M^{4j}\vert \lambda \vert^{\epsilon}\sigma^2 (x) dx )] \bigg\}
\end{eqnarray}
where $\chi$ is a function with compact support independent of $j$ and $\lambda$.

The small field region $S$ is the union of all the cubes for which the $\chi$ 
factor has been chosen. The complement, called the large field region $L$, 
is decomposed  as the union of connected pieces $L_k$.
Each such connected large field region has a small probabilistic factor for each of its cube 
using e.g. some standard Tchebycheff inequality.

The field is decomposed according to its localization
as  $\sigma = \sigma_S + \sum_k\sigma_{L_k}$.
Then the resolvent $C_j (\sigma, x, y )$ 
is simply bounded in norm if $x$ and $y$ belong to the same $L_k$
region because the decay is provided by the probabilistic factor associated to $L_k$.

The $\sigma_S$ piece is expanded according to resolvent formulas such as  
\begin{eqnarray}
C_j(\sigma_S, x, y)   = C_j(x, y) -  i \sqrt{\lambda} 
\int  dz C_j(x, z) \sigma_S (z)  C_j(\sigma_S, z, y),
\end{eqnarray}
which can be iterated to infinity because the $\sigma_S$ field is not integrated with 
the Gaussian measure but bounded with the help of the small field conditions.

Then inside each connected large field region $L_k$ the resolvent $C_j (\sigma_{L_k}, x, y )$ 
is simply bounded in norm. The decay is provided by the probabilistic factor 
associated to $L_k$. Between different connected large field regions, the decay
is provided by the small field resolvent expansion.

However one advantage of the loop expansion presented in this paper is to avoid the need
of any lattice of cubes for cluster/Mayer expansions. If possible, it seems better to us
to avoid reintroducing a lattice of cubes in such a small/large field analysis.

\subsection{Multiscale Expansions}

The result presented in this paper for a single scale model should be extended
to a multiscale analysis. This means that every loop-vertex or resolvent should carry a
scale index $j$ which represents the $lowest$ scale which appears in that loop
or resolvent. Then we know that the forest formula used in this paper should 
be replaced by a so-called ``jungle" formula \cite{AR1} in which links are built preferentially between loop vertices and resolvents
of highest possible index.

This jungle formula has to be completed by a ``vertical expansion" which tests whether connected contributions of higher scales have less or more than four external legs of lower scales, see e.g. \cite{AR3}.
A renormalization expansion then extracts the local parts of the corresponding two and four point contributions
and resums them into effective couplings. In this way
it should be possible to finally complete the program \cite{AR3} of 
a Bosonic renormalization-group-resummed expansion 
whose pieces are defined through totally explicit formulas without using any induction. 
Indeed the missing ingredient in \cite{AR3}, namely
an explicit formula to insert \emph{Mayer expansions} between each cluster expansion, 
would be totally avoided. The new multiscale expansion 
would indeed not require any cluster nor Mayer expansion at any stage.

The expansion would be completed by auxiliary resolvent expansions, either
with integration by parts in the manner of section 
\ref{iterresol} or with a small/large field analysis as in subsection \ref{largefield} above. 
This is necessary to establish scaled spatial decay, which in turn is crucial to prove that 
the renormalized two and four point contributions are small. 
But these new auxiliary expansions shall be used
only to prove the desired bounds, not to define the expansion itself.

\subsection{Vector Models}

The method presented here is especially suited to the treatment of 
large $N$ vector models. Indeed we can decompose a vector $\phi^4$ interaction
with an intermediate scalar field as in (\ref{intermediate})
but in such a way that the flow of vector indices occurs
within the loop-vertices. Every loop vertex simply carries therefore a global $N$ factor
where $N$ is the number of colors. Hence we expect that the loop expansion presented here
is the right tool to glue different regimes of the renormalization group
governed respectively e.g. in the ultraviolet regime by a small coupling expansion and 
in the infrared regime by a ``non-perturbative" large $N$ expansion of the vector type. 
This gluing problem occurs in many different physical contexts, from mass generation of the two-dimensional  Gross-Neveu \cite{KMR} or non-linear $\sigma$-model \cite{K} to the BCS theory of supraconductivity \cite{FMRT}. These gluing problems have been considered until now too complicated in practice for a rigorous constructive analysis.

\subsection{Matrix models and $\phi^{\star 4}_4$}

The loop expansion is also suited for the treatment of large $N$ matrix 
models and was in fact found for this reason \cite{Riv3}. Our first goal 
is to apply it to the full construction of non-commutative $\phi^{\star 4}_4$ 
\cite{GW}, either in the so-called matrix base \cite{GW2,RVW} or in 
direct space \cite{GMRV}. 

One needs again to develop for that purpose the multiscale version of the expansion
and the resolvent bounds analogs to section \ref{iterresol} or subsection \ref{largefield}
above. Indeed neither the matrix propagator nor the Mehler $x$ space propagator 
are diagonal in the corresponding representations/footnote{There is an interesting exception: the matrix propagator 
of $\phi^{\star 4}_4$ becomes diagonal in the matrix base at the very special 
ultraviolet fixed point where $\Omega$, the Grosse-Wulkenhaar
parameter, is 1, Of course the general 
non-diagonal case has to be treated.}. 

Ultimately we hope that better understanding the non-commutative models
of the matrix or quasi-matrix type should be useful in many areas of physics, from
physics beyond the standard model \cite{CCM,Co,DN} 
to more down to earth physics such as quark confinement
\cite{Hoo} or the quantum Hall effect \cite{Poly}.

\end{document}